# Corotating Magnetic Reconnection Site in Saturn's Magnetosphere

Z. H. Yao[1,2], A. J. Coates[1,3], L. C. Ray[1,4], I. J. Rae[1], D. Grodent[2], G. H. Jones[1,3], M. K. Dougherty[5], C. J. Owen[1], R. L. Guo[6], W. R. Dunn[1,3], A. Radioti[2], Z. Y. Pu[7], G. R. Lewis[1], J. H. Waite[8], and J.-C. Gérard[2]

[1] UCL Mullard Space Science Laboratory, Dorking RH5 6NT, UK; z.yao@ucl.ac.uk
[2] Laboratoire de Physique Atmosphérique et Planétaire, STAR institute, Université de Liège, B-4000 Liège, Belgium
[3] Centre for Planetary Sciences at UCL/Birkbeck, Gower Street, London WC1E 6BT, UK
[4] Department of Physics, Lancaster University, Lancaster, UK
[5] Imperial College of Science, Technology and Medicine, Space and Atmospheric Physics Group, Department of Physics, London SW7 2BW, UK
[6] Key Laboratory of Earth and Planetary Physics, Institute of Geology and Geophysics, Chinese Academy of Sciences, Beijing, China
[7] School of Earth and Space Sciences, Peking University, Beijing, China
[8] Southwest Research Institute, San Antonio, TX, USA

## Abstract

Using measurements from the *Cassini* spacecraft in Saturn's magnetosphere, we propose a 3D physical picture of a corotating reconnection site, which can only be driven by an internally generated source. Our results demonstrate that the corotating magnetic reconnection can drive an expansion of the current sheet in Saturn's magnetosphere and, consequently, can produce Fermi acceleration of electrons. This reconnection site lasted for longer than one of Saturn's rotation period. The long-lasting and corotating natures of the magnetic reconnection site at Saturn suggest fundamentally different roles of magnetic reconnection in driving magnetospheric dynamics (e.g., the auroral precipitation) from the Earth. Our corotating reconnection picture could also potentially shed light on the fast rotating magnetized plasma environments in the solar system and beyond.

## 1. Introduction

The terrestrial magnetospheres receive energy from the Sun via dayside magnetopause reconnection and occasionally explosively release this energy, causing strong perturbations within their nightside magnetospheres, ionospheres, and atmospheres. Electrical currents that flow along the magnetic field couple the magnetosphere and the ionosphere, disturbing the polar geomagnetic field and powering the aurorae (Akasofu 1964; McPherron et al. 1973). Magnetospheric perturbations and auroral intensifications also exist at other planets, including the giant planets (i.e., Saturn and Jupiter; Clarke et al. 2005). Unlike at Earth, internally generated plasma sources (moons and rings) at the giant planets play an important role in driving dynamics in planetary magnetotails (Kivelson & Southwood 2005; Kronberg et al. 2007). The centrifugal forces caused by the rotation of the giant planets are usually considered to be the main mechanism in driving internal plasma dynamics in a process referred to as the Vasyliunas cycle (Vasyliunas 1983). The solar wind controlled, externally driven large-scale magnetospheric circulation of energy defined for Earth's magnetosphere is well known as the Dungey cycle (Dungey 1961).

The magnetic field in Saturn's magnetosphere consists of two contributions, i.e., the magnetospheric currents and the planetary dipole field. The magnetospheric currents can from time to time divert into the ionosphere and cause auroral intensifications (Mitchell et al. 2005). During this process, the magnetic field in the magnetosphere would naturally become more dipolar. This process is also known as magnetic dipolarization, as the magnetospheric current contributes less to the measured magnetic field so that the dipole magnetic field can have a relatively greater contribution. At Saturn and Jupiter, the internally driven Vasyliūnas cycle strongly contributes to the dynamics of their rotationally dominated magnetospheres. The Vasyliūnas cycle is suggested to take place on closed field lines, with the location of the reconnection line increasing in distance from the planet from dusk to dawn (Vasyliunas 1983; Kivelson & Southwood 2005; Delamere et al. 2015). In situ evidence directly linking the Vasyliūnas cycle to global magnetospheric dynamics is rare, although the quasi-periodicity of the particle and magnetic field perturbations at Jupiter implies that the planet's rotation may participate in planetary energy reloading (Kronberg et al. 2005). Statistical studies of reconnection events at Jupiter (Vogt et al. 2014) and Saturn (Jackman et al. 2014) suggest that the reconnection line orientation may oppose that predicted by simulations and theoretical models of the Vasyliūnas cycle. So far, it is still unclear how the Vasyliūnas cycle drives the dynamics of giant magnetospheres. Energetic particle injections in the inner Saturnian magnetosphere have been shown to be related to radial transport, driven by the centrifugal interchange instability (Hill et al. 2005), a separate physical process from the Vasyliūnas cycle. Here, we report the first direct observation of an internally driven corotating magnetic reconnection event associated with explosive energy release (EER) and the subsequent reloading of magnetic energy at Saturn. Regarding the different environments of Saturn and Earth, we hereby define an Earth substorm-like EER as a process that includes current sheet north–south expansion (i.e., thickening), particle acceleration, and magnetic dipolarization.

## 2. Cassini Observations of the Quasi-steady Reconnection Site

Figures 1(a)–(c) show the magnetic field data from the *Cassini* magnetometer (Dougherty et al. 2004) in Kronographic Radial–Theta–Phi coordinates during 2006 September 20 and 21. This is a Saturn-centered coordinate. Namely, the radial vector (**r**) is

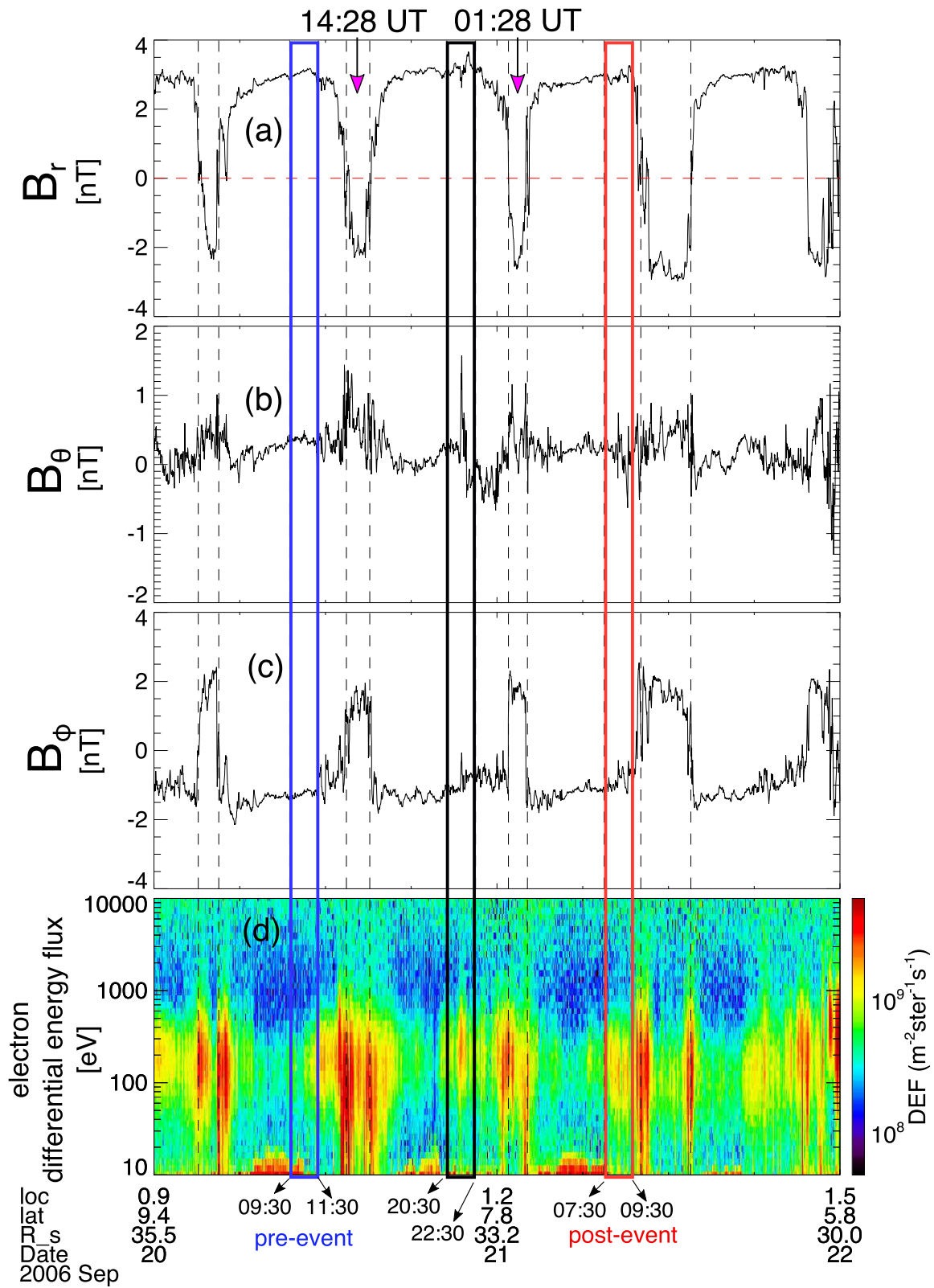

**Figure 1.** ((a)–(c)) Overview of one-minute resolution magnetic field $B_r$, $B_\theta$, and $B_\Phi$ components from the *Cassini* magnetometer (MAG) in Kronographic Radial–Theta–Phi coordinates, and (d) electron differential energy flux from the *Cassini* electron spectrometer (CAPS-ELS). The black rectangle selects the time period from September 20 20:30 UT to 22:30 UT including the dipolarization event, the blue rectangle selects the period from September 20 09:30 UT to 11:30 UT, and the red rectangle selects the period from September 21 07:30 UT to 09:30 UT. The two pink arrows in panel (a) indicate the two minimums of $B_r$; the separation represents the rotation period, i.e., 11 hr.

directed from the center of Saturn to the spacecraft, the azimuthal component ($\phi$) is parallel to the equator and positive in the direction of corotation, and the "southward" $\theta$ component completes the right-hand set ($\theta = r \times \phi$). During this period, *Cassini* was located post-midnight between 0.9 and 1.5 Saturn local time, at latitudes from $\sim 9.4°$N to $5.8°$N north of Saturn's equatorial plane and at radii from $R \sim 30$ to $35\ R_S$ (1 $R_S = 60268$ km). Periodic current sheet crossings caused by Saturn's rotation are identified by the reversals of $B_r$ and $B_\Phi$ ($B_r$, $B_\Phi \sim 0$ are marked by the vertical dashed lines). These current sheet crossings are accompanied by enhancements of electron flux (all anodes averaged flux; Figure 1(d)) with energies of tens to hundreds of eV. The black rectangle indicates a significant $B_\theta$ component magnetic field spike at around 21:30 UT, accompanied by an electron flux enhancement, which represents magnetic dipolarization (Yao et al. 2017a).

We define the measurements made one Saturn rotation period before the event as pre-event (a baseline) and from one Saturn rotation period after the events as post-event. Saturn's magnetospheric rotation period is time variable (Gurnett et al. 2009). We determine here a period from the large-scale current sheet crossing, which can better represent the magnetospheric rotation. As indicated by the pink arrows in Figure 1(a), the time difference between the pre-event and dipolarization event is 11 hr (separation between the two pink arrows). Moreover, the major change of magnetic field components $B_\theta$ and $B_\Phi$ between the dipolarization event and pre-event is highly consistent. It is thus reasonable to adopt 11 hr as the magnetospheric rotation period during our event.

In Figure 2, we compare the measurements within the dipolarization event between 20:30 UT and 22:30 UT, pre-event (between 20:30 UT − 11 hr and 22:30 UT − 11 hr), and post-event (between 20:30 UT + 11 hr and 22:30 UT + 11 hr). The magnetic field variations shown in Figure 2 are generally consistent among the three periods before $T_0 + 1$ hr, suggesting that the three periods of measurements were made in the same region of the rotating magnetosphere when the planet rotated twice. Figure 2(d) shows the magnetic elevation angle that is defined $\tan^{-1}\left(\frac{|B_\theta|}{\sqrt{(B_r^2 + B_\Phi^2)}}\right)$. This angle indicates the morphological change of the magnetic field line from tail-like to the dipolar at Saturn (Arridge et al. 2009), and a similar indicator has been widely applied in identifying terrestrial magnetic dipolarization (Shiokawa et al. 2005). This indicator is enhanced from less than 5° to ∼30° during the dipolarization event and to ∼15° during post-event. In the pre-event period, the magnetic field was not significantly perturbed, with a dominant component $B_r \sim 3$ nT, and a relatively low electron flux, suggesting *Cassini* was in the outer plasma sheet or lobe region (Arridge et al. 2009). One rotation later, a significant magnetic dipolarization signature ($B_r$ decrease and $B_\theta$ increase) was observed around 21:30 UT, followed by $B_\theta$ decreasing to negative. Meanwhile, the electron flux was enhanced during the magnetic dipolarization. The $B_r$ decrease, $B_\theta$ increase, and electron flux enhancements are typical features of current sheet expansion and magnetic dipolarization in Earth's magnetotail (Lui 1996; Angelopoulos et al. 2008), so this event satisfies our definition of an Earth substorm-like EER event (Yao et al. 2012, 2017a).

In addition to the previous reconnection evidence, the clear $B_\Phi$ decrease that accompanies the $B_\theta$ increase is likely driven by the magnetic tension force when reconnection suddenly reduces the length of the magnetic field line. The $B_\Phi$ decrease may also relate to Saturn's supercorotational return flow phenomenon (Masters et al. 2011). In the post-event measurements, a $B_\theta$ enhancement and $B_\Phi$ decrease are also observed. Obviously, the variations in post-event measurements are similar to the dipolarization event measurements, although with a smaller amplitude. We thus conclude that the post-event state is developed from the dipolarization condition after a significant magnetic energy reloading process. We use Figure 2(h) to illustrate the evolution of the magnetic field configuration from pre-event to post-event. Pre-event, the magnetic field was highly stretched (blue) and *Cassini* was in the lobe (or outer plasma sheet) region. After the Earth substorm-like EER event, the magnetic field is dipolarized (black), and *Cassini*, relatively, moves into the plasma sheet. Thereafter, the reloading process stretches the magnetic field to approach the pre-event configuration (red).

## 3. Modeling of Fermi Acceleration for a Dipolarization at Saturn

During a magnetic dipolarization process, a shrinking length of magnetic field line is naturally expected (Wu et al. 2006; Yao et al. 2017b), which causes Fermi acceleration when the second adiabatic invariant is conserved. Considering the short bouncing period and small gyroradius of electrons, it is reasonable to assume the conservation of the second adiabatic invariant for electrons in the magnetosphere. Therefore, parallel Fermi acceleration of electrons is expected during a magnetic dipolarization (Wu et al. 2006; Birn et al. 2013). As shown in Figure 3(a), modified from the picture presented in the plasma textbook *Physics of Space Plasmas: An Introduction* by George Parks (Parks 1991), particles moving between two wall mirror points would gain energy for each collision with a wall if the walls approach each other. From the second adiabatic invariant, we know that the parallel velocity $V_{//}$ satisfies the relation $V_{//}*D=$ constant (Equation (4.127) in Parks 1991), where $D$ is the distance between two mirror points (see Figure 3(a)). So the reduction of $D$ would naturally lead to a parallel pressure increase, i.e., the Fermi acceleration. In Saturn's magnetosphere, particles bounce between mirror points in the northern and southern hemisphere ionospheres with the length of the magnetic field line equivalent to the distance "$D$" in Parks's schematic. During magnetic dipolarization, the magnetic field line shrinks, naturally causing Fermi acceleration.

The degree of Fermi acceleration depends on how much the magnetic field length has been changed, i.e., $V_{//} \sim 1/D$. We roughly estimate the efficiency of Fermi acceleration at Saturn from a simple magnetic field model (Wu et al. 2006) as shown in Figure 3(b). The red curve represents the dipolarized field, and the black curve represents a stretched magnetic field before dipolarization. Before dipolarization, *Cassini* was in the outer plasma sheet, suggesting that the magnetic field was near the open–closed field line. As given in a classic plasma textbook (Baumjohann & Treumann 1997), the change of arc element along a dipolar magnetic field is given by $\frac{ds}{d\lambda} = r_{eq}\cos\lambda(1 + 3\sin^2\theta)^{1/2}$ (Equation (3.7) in their book). We thus obtain the length of the red curve magnetic field line ($S'$) in our schematic plot to be 91 $R_S$. The length of the black curve is roughly $S = 91\ R_S + 2*L2$. The exact length of L2 is very difficult to determine, However, from previous literature, *Cassini* has often measured the tailward reconnection site within 60 $R_S$, suggesting that the open–closed magnetic field

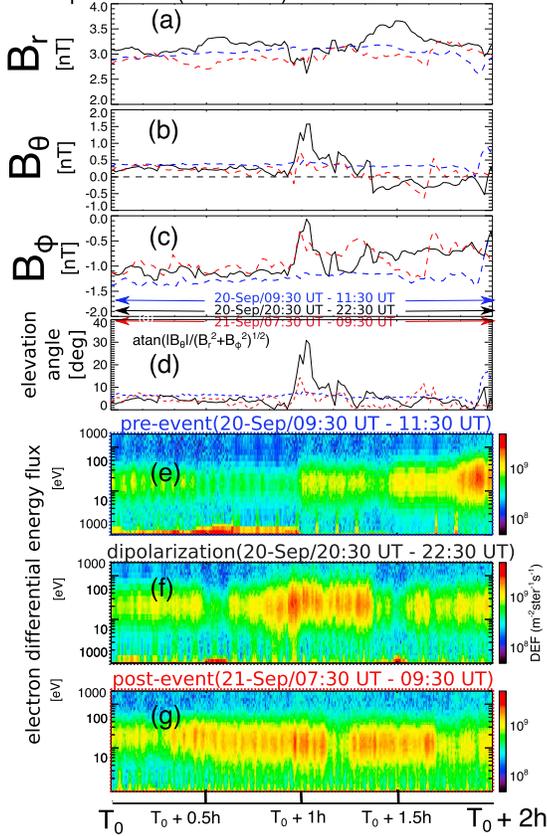

A comparison of shifted pre-event (−11 hours), dipolarization, and post-event (+11 hours) measurements over 2 hour periods line is very often within 60 $R_S$ (Yao 2017). We adopt here a range of L2 ∼ 10–30 $R_S$ to calculate the effect of Fermi acceleration. Although this is an estimate, it does not affect our major conclusion that the measured acceleration is likely Fermi acceleration instead of reconnection acceleration. We therefore adopt $S = 91 \, R_S + 2^* L2 = 111$–151 $R_S$, so $S/S' \sim 1.2$–1.6 and $\frac{E'_{\parallel}}{E_{\parallel}} = \left(\frac{v'_{\parallel}}{v_{\parallel}}\right)^2 = \left(\frac{S}{S'}\right)^2 \sim 1.4$–2.7.

**Figure 2.** Comparison of the pre-event, during-event, and post-event in situ data with proposed magnetic field topology. One-minute resolution magnetic field ((a)–(c)), magnetic elevation angle ($\tan^{-1}(\frac{|B_\theta|}{\sqrt{(B_r^2 + B_\phi^2)}})$) (d), and electron differential energy flux ((e)–(g)) during the unperturbed period (dashed blue magnetic field and panel (e) for electron flux; the blue rectangle in Figure 1), dipolarization period (solid black magnetic field and panel (f) for electron flux; the black rectangle in Figure 1), and post-dipolarization (dashed red magnetic field and panel (g) for electron flux) periods (the red rectangle in Figure 1). The unperturbed measurements are the in situ *Cassini* measurements shifted forward by 11 hr, and the post-dipolarization measurements are the in situ *Cassini* measurements shifted backward by 11 hr. (h) Geometry of the magnetic field topology and schematic of *Cassini*'s location (star) relative to the current sheet. The blue magnetic field line illustrates the stretched magnetic field, which represents a thin current sheet condition. The black and red curves represent the dipolarized and post-dipolarization magnetic field lines, respectively.

Figure 4 shows the differential electron flux (parallel to **B**) from the dipolarization region and from the background region with the estimated change in the parallel energy resulting from Fermi acceleration. We model the Fermi acceleration of two possible background populations: the pre-dipolarization distribution (2006 September 20/21:10 UT; green) and the nearest current sheet distribution (2006 September 21/00:47 UT; red). In our event, the peak energy of background populations shifted from 130 to 300 eV, $\frac{E'_{\parallel}}{E_{\parallel}} \sim 2.3$, which is a reasonable value for our modeling calculation of Fermi acceleration (i.e., 1.4–2.7). The accelerated distributions are in good agreement with that of the dipolarization population, although slight differences exist, which could be due to some non-adiabatic acceleration, or measurements uncertainties. The general agreement between our modeled acceleration populations and the in situ measurements demonstrates that Fermi acceleration was the major mechanism in accelerating the electrons during this dipolarization event.

In the magnetospheric environment, electrons bounce back and forth between two mirror points. The lower pitch angle (more aligned to the magnetic field) electrons can reach closer to the planet, and larger pitch angle electrons (more perpendicular to the magnetic field) will be confined near the equator. Since all particles travel through the magnetic equator, the measurements from the equator would be ideal to investigate an energization process. The betatron accelerated electrons would have large pitch angles, which are confined near the equator and cannot be well detected by the spacecraft in the outer plasma sheet. From Liouville's theorem, it is impossible to derive a distribution on the equator from the measurements away from the equator. In the region where the *Cassini* spacecraft was located, the betatron acceleration is not efficient. This is because that betatron acceleration depends on the change of the magnetic strength, not only on the $\boldsymbol{B_\theta}$ component. Prior to dipolarization, the magnetic strength $B \sim 3.4$ nT and during the dipolarization $B \sim 3.1$ nT, so that the perpendicular energization from the betatron acceleration was not expected at the spacecraft location as the magnetic strength was not enhanced. The perpendicular acceleration of electron population can only come from the central plasma sheet, and the final energies would depend on the ratio of dipolarization and initial magnetic strength at the equator, which cannot, however, be assessed by the *Cassini* in situ measurements in this research. By comparing the distributions between pre-dipolarization and dipolarization regions (not shown in this Letter), we found that the energy of the peak flux shifted up less for higher pitch angles, and there was almost no change for the perpendicular population. This suggests that even though there was strong betatron acceleration in the central current sheet, the population must have been strongly confined in the inner plasma sheet. The betatron acceleration associated with the dipolarization is thus not discussed in this study.

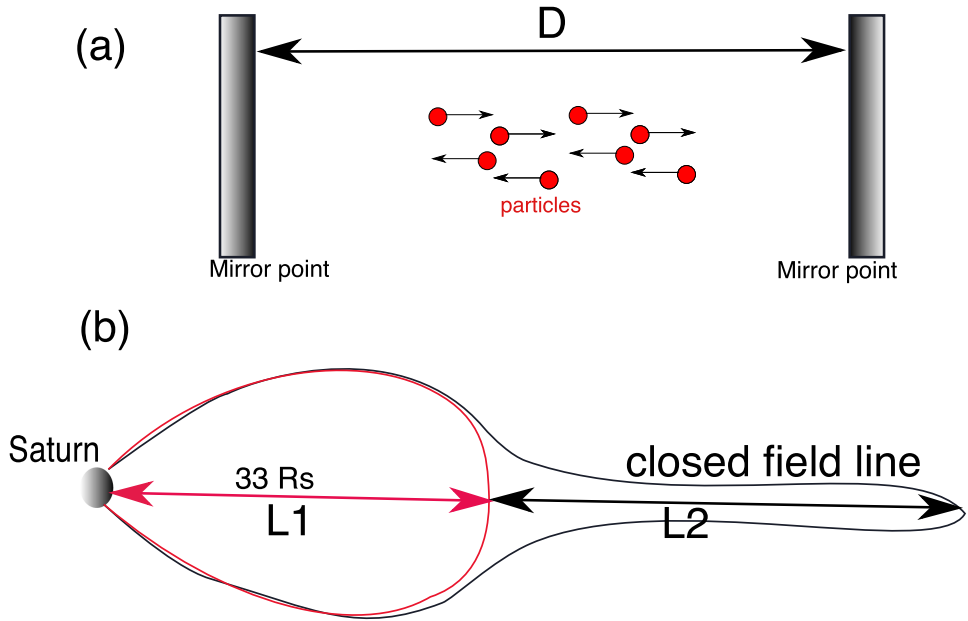

**Figure 3.** (a) Particles bounce between two mirror points. (b) Schematic of the magnetic field configuration associated with dipolarization. The black curve represents a magnetic field line before dipolarization, and the red curve represents the magnetic field line after dipolarization. L1 is the equatorial distance of the red curve to Saturn, and L1+L2 is the equatorial distance of the black curve to Saturn.

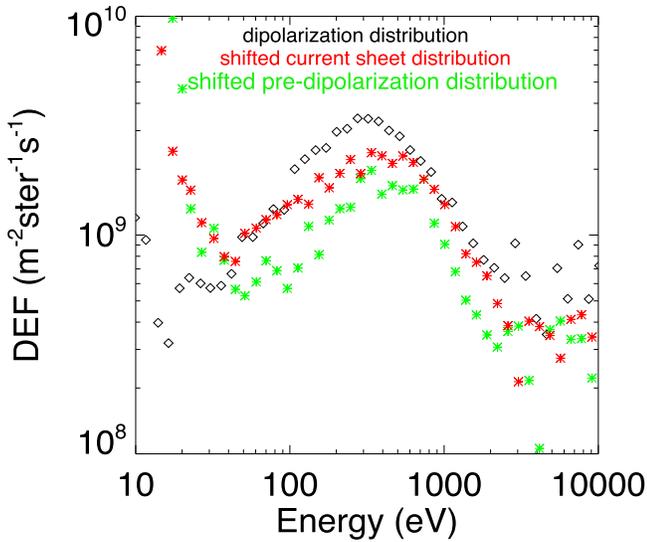

**Figure 4.** Differential electron distribution (parallel to **B**) from the dipolarization region (black) and from the background region with an estimated change in the parallel energy resulting from a Fermi acceleration (shifted by a factor of 2.3 in energy). We choose the pre-dipolarization distribution (2006 September 20/21:10 UT; green) and the nearest current sheet distribution (2006 September 21/00:47 UT; red) as two possible background populations.

However, the *Cassini* measurements in the outer plasma sheet can well detect the Fermi acceleration population. In this study, the measured parallel electron population have pitch angle $\alpha < 20°$, i.e., $\frac{V_\perp}{V_\parallel} = \tan \alpha < 0.36$. Thus, $W_\perp / W_\parallel = (V_\perp / V_\parallel)^2 = (\tan \alpha)^2 < 13\%$. The energy of measured electron population is mostly carried by the parallel energy ($>87\%$). The energy would convert from the perpendicular component to the parallel component if we trace the population to the equator plane. However, this conversion is only less than 13%, much less than the Fermi acceleration discussed in this study, i.e., $\frac{E_\parallel'}{E_\parallel} \sim 2.3$. Therefore, it is reasonable to investigate the Fermi acceleration process with the measurements from the plasma sheet boundary layer.

To explain the $B_\theta$ positive to negative feature in the dipolarization event and the post-event, we propose a corotating magnetic reconnection picture, as presented in Figure 5. As the magnetic reconnection region corotates with Saturn (from red to blue), *Cassini*'s relative position moves from Saturnward (Figure 5(a)) to tailward (Figure 5(b)), consequently observing the reversal of $B_\theta$, together with the decrease of $B_\Phi$ (the two panels below the cartoons). Figure 5(c) shows the equatorial projection of the reconnection X-line from Figures 5(a) to (b). In previous literature, the $B_\theta$ reversal is usually described as a tailward plasmoid (Vogt et al. 2014), which is based on a 2D picture, i.e., not including azimuthal variation. However, the retreat of a 2D plasmoid structure cannot well explain a similar $B_\theta$ reversal signature during post-event (the red curves in Figure 5), which can be well explained by our corotating magnetic reconnection picture. It is interesting to note that hot electrons are observed at the tailward reconnection site region, which is evidence that the reconnection operated on closed field lines so that both the planetward and tailward regions of the reconnection site were not opened to the interplanetary magnetic field.

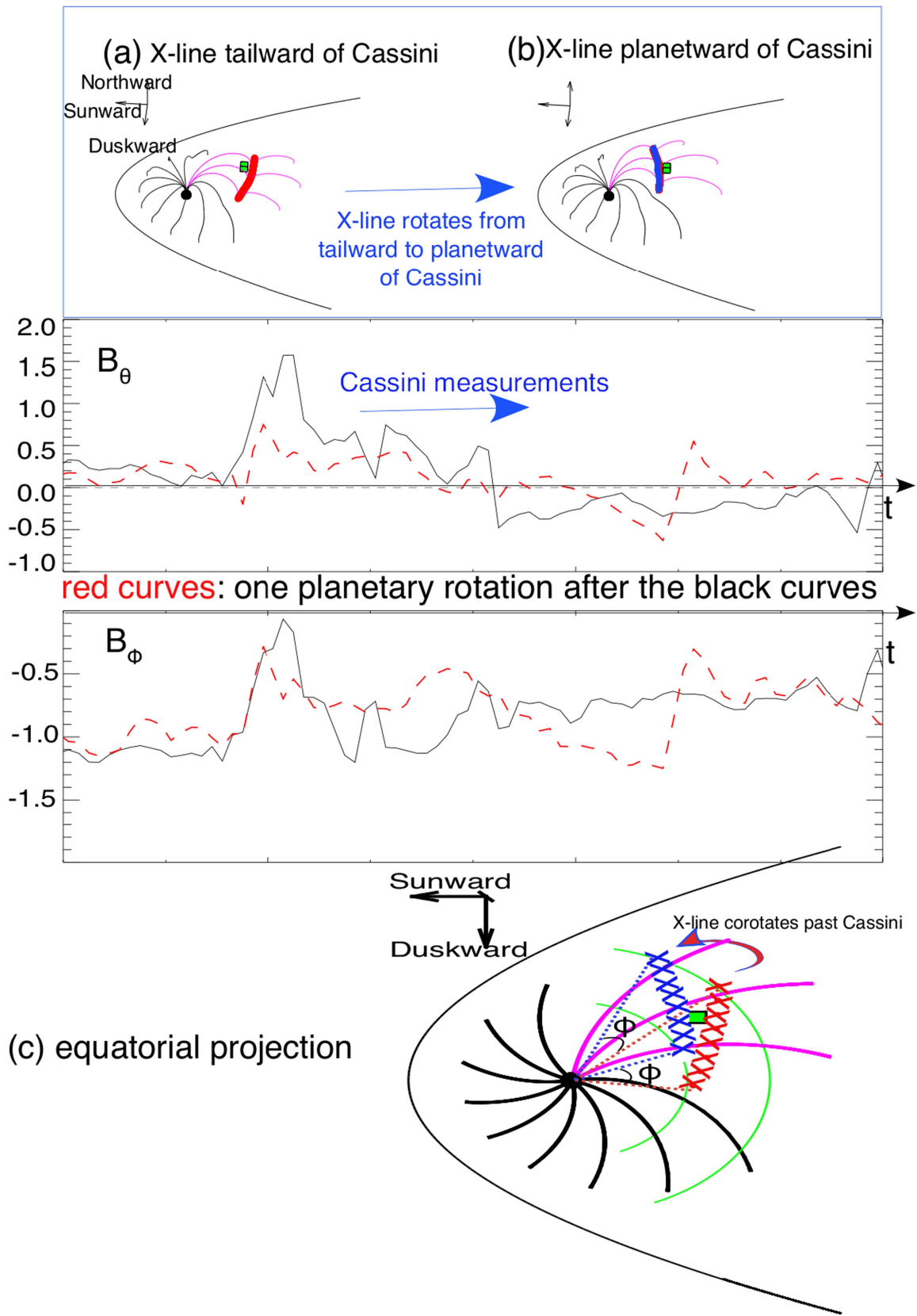

**Figure 5.** Illustration of the corotating magnetic reconnection picture and connection to in situ measurements. (a) A sketch showing the reconnection line (red curve) and the location of the spacecraft (green rectangle) Saturnward of the reconnection line (red). (b) After a small rotation from (a), the spacecraft was tailward of the reconnection line (blue). (c) The equatorial projection of the corotating picture from the X-line in red (a) to the X-line in blue (b).

The azimuthal scale of the reconnection line in our event is about 2 MLT (or $\sim 1.2 \times 10^6$ km), corresponding to the $\sim 1$ hr duration between 21:30:00 UT and 22:30:00 UT. We would like to point out that corotating auroral features have been previously observed to break up with a 3 MLT azimuthal scale, although observations there have not been discussed in the context of a corotating reconnection picture (Radioti et al. 2016). A rotating magnetic reconnection region could generate such a corotating aurora breakup region in strong support of our picture.

## 4. Conclusion

The agreement between our corotating reconnection picture and the *Cassini* in situ measurements is striking, indicating that the reversal of $B_\theta$ cannot be fully explained by a 2D magnetic reconnection retreat picture, which can either be driven by the solar wind or an internal source. Previous literature has reported a reconnection event at Saturn lasting for up to 19 hr, and the event has been suggested to be likely driven by solar wind (Arridge et al. 2016). We must emphasize that if this long-lasting reconnection is driven by the solar wind, the reconnection site should not corotate with the planet. However, our corotating reconnection picture can only be driven by an internal source, since the tailward region of solar wind driven reconnection is disconnected from Saturn's magnetosphere, and thus cannot be observed after one rotation. The internally driven reconnection process is also confirmed by the fact that hot electrons were measured in the tailward reconnection site region. The reloading process of magnetic energy after an Earth substorm-like EER is longer than 11 hr, and hence is significantly longer than the 3–4 hr typically measured at Earth (Akasofu 1964). The $B_\theta$ positive to negative signature exists during two rotations, indicating that it corresponds to a quasi-steady structure. As such, we suggest that this bipolar signature is a rotating spatial structure. Therefore, the peak of $B_\theta$ may not represent a formation of magnetic dipolarization, but just a signature that a previously formed dipolarization encountered by the spacecraft during its rotation. Thus, the measured accelerated electrons were trapped in the $B_\theta$ peak structure. A similar mechanism to that of the energetic particle trapped around the localized magnetic field peak is found in a terrestrial dipolarizing flux bundle from simulations (Gabrielse et al. 2017). This acceleration process should be clearly distinguished from the induced electric field acceleration of the ambient plasma due to the motion of a dipolarization front (Runov et al. 2015; Gabrielse et al. 2016). We would like to point out that the corotating reconnection site would naturally suggest the existence of reconnection in the dayside magnetosphere, which has been supported by indirect statistical evidence (Delamere et al. 2015).

Distinguishing temporal evolution from spatial structures is a major difficulty of single-spacecraft in situ measurements, limiting the physical interpretation of data. We propose that by comparing measurements from subsequent rotations, when the location of *Cassini* can be approximated as nearly constant, temporal and spatial variations can be decoupled. This technique can be widely applied to rotating giant planet systems. This work sets an exciting example on how taking advantage of comparing measurements over subsequent planetary rotations can decouple temporal and spatial variations, overcoming the limitations of single-spacecraft in situ data. We expect a similar process to exist in Jupiter's magnetosphere, which generates the most powerful aurora in the solar system. The NASA/*JUNO* spacecraft, currently in a polar orbit at Jupiter, can directly examine our prediction (Bagenal et al. 2014).

Z.Y., I.J.R., C.J.O., A.J.C., W.R.D., and G.H.J. are supported by a UK Science and Technology Facilities Council (STFC) grant (ST/L005638/1) at UCL/MSSL. Z.Y. is a Marie-Curie COFUND postdoctoral fellow at the University of Liege. Co-funded by the European Union. R.L.G is supported by the National Science Foundation of China (41525016, 41404117). A.R. is funded by the Belgian Fund for Scientific Research (FNRS). Z.Y. very much appreciates useful and fresh discussion from Jennifer Y. H. Chan at MSSL/UCL and Peter Delamere at University of Alaska Fairbanks. *Cassini* operations are supported by NASA (managed by the Jet Propulsion Laboratory) and ESA. The data presented in this Letter are available from the NASA Planetary Data System https://pds-ppi.igpp.ucla.edu/.

### ORCID iDs

A. J. Coates ● https://orcid.org/0000-0002-6185-3125
L. C. Ray ● https://orcid.org/0000-0003-3727-602X
I. J. Rae ● https://orcid.org/0000-0002-2637-4786
D. Grodent ● https://orcid.org/0000-0002-9938-4707
M. K. Dougherty ● https://orcid.org/0000-0002-9658-8085
C. J. Owen ● https://orcid.org/0000-0002-5982-4667
R. L. Guo ● https://orcid.org/0000-0002-7125-0942
W. R. Dunn ● https://orcid.org/0000-0002-0383-6917

## References

Akasofu, S. I. 1964, P&SS, 12, 273
Angelopoulos, V., McFadden, J. P., Larson, D., et al. 2008, Sci, 321, 931
Arridge, C. S., Eastwood, J. P., Jackman, C. M., et al. 2016, NatPh, 12, 268
Arridge, C. S., McAndrews, H. J., Jackman, C. M., et al. 2009, P&SS, 57, 2032
Bagenal, F., Adriani, A., Allegrini, F., et al. 2014, SSRv, in press
Baumjohann, W., & Treumann, R. A. 1997, Basic Space Plasma Physics (Singapore: World Scientific)
Birn, J., Hesse, M., Nakamura, R., & Zaharia, S. 2013, JGRA, 118, 1960
Clarke, J., Gerard, J.-C., Grodent, D., et al. 2005, Natur, 433, 717
Delamere, P., Otto, A., Ma, X., Bagenal, F., & Wilson, R. 2015, JGRA, 120, 4229
Dougherty, M. K., Kellock, S., Southwood, D. J., et al. 2004, SSRv, 114, 331
Dungey, J. W. 1961, PhRvL, 6, 47
Gabrielse, C., Angelopoulos, V., Harris, C., et al. 2017, JGRA, 122, 5059
Gabrielse, C., Harris, C., Angelopoulos, V., Artemyev, A., & Runov, A. 2016, JGRA, 121, 9560
Gurnett, D., Lecacheux, A., Kurth, W. S., et al. 2009, GeoRL, 36, L16102
Hill, T., Rymer, A. M., Burch, J. L., et al. 2005, GeoRL, 32, L14S10
Jackman, C., Slavin, J. A., Kivelson, M. G., et al. 2014, JGRA, 119, 5465
Kivelson, M., & Southwood, D. 2005, JGRA, 110, A12209
Kronberg, E., Glassmeier, K. H., Woch, J., et al. 2007, JGRA, 112, A05203
Kronberg, E., Woch, J., Krupp, N., et al. 2005, JGRA, 110, A03211
Lui, A. T. Y. 1996, JGRA, 101, 13067
Masters, A., Thomsen, M., Badman, S., et al. 2011, JGRA, 116, A10212
McPherron, R., Russell, C., Kivelson, M., & Coleman, P., Jr. 1973, RaSc, 8, 1059
Mitchell, D., Brandt, P. C., Roelof, E. C., et al. 2005, GeoRL, 32, L20S01
Parks, G. K. 1991, (Redwood City, CA: Addison-Wesley), 547
Radioti, A., Grodent, D., Jia, X., et al. 2016, Icar, 263, 75
Runov, A., Angelopoulos, V., Gabrielse, C., et al. 2015, JGRA, 120, 4369
Shiokawa, K., Shinohara, I., Mukai, T., Hayakawa, H., & Cheng, C. 2005, JGRA, 110, A05212
Vasyliunas, V. M. 1983, in Physics of the Jovian Magnetosphere, ed. A. J. Dessler (Cambridge: Cambridge Univ. Press), 395
Vogt, M. F., Jackman, C. M., Slavin, J. A., et al. 2014, JGRA, 119, 821
Wu, P., Fritz, T., Larvaud, B., & Lucek, E. 2006, GeoRL, 33, L17101
Yao, Z., Grodent, D., Ray, L. C., et al. 2017a, JGRA, 122, 4348
Yao, Z. H. 2017, Earth Planet. Phys., 1, 53
Yao, Z. H., Pu, Z. Y., Fu, S. Y., et al. 2012, GeoRL, 39, L13102
Yao, Z., Rae, I. J., Lui, A. T. Y., et al. 2017b, JGRA, in press